\documentclass[a4paper,fleqn,usenatbib]{mnras}

\usepackage{newtxtext,newtxmath}
\usepackage{mathptmx}
\usepackage{txfonts}

\usepackage[T1]{fontenc}
\usepackage{ae,aecompl}

\usepackage{graphicx}	
\usepackage{amsmath}	
\usepackage{amssymb}	
\usepackage[utf8]{inputenc}
\usepackage[export]{adjustbox}
\usepackage{breqn}
\usepackage{subfig}
\usepackage{listings}
\usepackage{multirow, makecell}
\usepackage{hyperref}

\title[Binary Asteroseismic Modelling]{Binary Asteroseismic Modelling: isochrone-cloud methodology and application to {\it Kepler} gravity mode pulsators}
\author[C. Johnston et al.]{
C. Johnston$^{1}$\thanks{e-mail: \href{mailto:colecampbell.johnston@kuleuven.be}{colecampbell.johnston@kuleuven.be}}
A. Tkachenko$^{1}$,
C. Aerts$^{1,2}$,
G. Molenberghs$^{3,4}$
D. M. Bowman$^{1}$,
\newauthor
M. G. Pedersen$^{1}$,
B. Buysschaert$^{1,5}$, 
P. I. P{\'a}pics$^{1}$
\\
$^{1}$Instituut voor Sterrenkunde, KU Leuven, Celestijnenlaan 200D, B-3001 Leuven, Belgium\\
$^{2}$Department of Astrophysics, IMAPP, Radboud University Nijmegen, P.O. Box 9010, 6500 GL Nijmegen, The Netherlands\\
$^{3}$I-BioStat, Universieit Hasselt, Martelarenlaan 42, B-3500 Hasselt, Belgium\\
$^{4}$I-BioStat, KU Leuven, Kapucijnenvoer 35, B-3000 Leuven, Belgium\\
$^{5}$LESIA, Observatoire de Paris, PSL Research University, CNRS, Sorbonne Universites, UPMC Univ. Paris 06, Univ. Paris Diderot,\\
Sorbonne Paris Cite, 5 place Jules Janssen, 92195, Meudon, France\\
}

\date{Accepted XXX. Received YYY; in original form ZZZ}

\pubyear{2018}

\begin{document}
\label{firstpage}
\pagerange{\pageref{firstpage}--\pageref{lastpage}}
\maketitle

\begin{abstract}
  The simultaneous presence of variability due to both pulsations and binarity
  is no rare phenomenon. Unfortunately, the complexities of dealing with even 
  one of these sources of variability individually means that the other
  signal is often treated as a nuisance and discarded. However, both types of
  variability offer means to probe fundamental stellar properties in robust
  ways through asteroseismic and binary modelling. We present an efficient
  methodology that includes both binary and asteroseismic information to estimate
  fundamental stellar properties based on a grid-based modelling approach. We report
  parameters for three gravity mode pulsating {\it Kepler} binaries 
  , such as mass, radius, age, as well the mass of the convective core and 
  location of the overshoot region. We discuss the presence of parameter degeneracies 
  and the way our methodology deals with them. We provide asteroseismically 
  calibrated isochrone-clouds to the community; these are a generalisation of 
  isochrones when allowing for different values of the core overshooting in 
  the two components of the binary.
\end{abstract}




\section{Introduction}

The mass of the helium core at the end of core hydrogen 
burning on the main-sequence (MS) is a pivotal quantity 
in stellar structure and evolution calculations, since it 
is regulated by the interior physics of MS stars and dictates the 
subsequent evolution of stars. Internal mixing 
processes modify the amount of nuclear fuel 
available for burning \citep{maeder2009,meynet2013},
 thus, are of paramount importance 
to stellar structure and evolution calculations. Having a 
fully mixed convective core on the MS, intermediate- and 
high-mass stars are particularly sensitive to internal 
mixing processes, making their calibration and implementation 
a high priority in modelling efforts. Several different 
physical mechanisms work either independently 
or in conjunction with one another to form an internal 
mixing profile, which cannot be derived from first principles 
\citep[][for a recent review]{salaris2017}. However, from  
those processes which contribute to near core mixing in 
intermediate- and high-mass stars, we identify two classes: 
i) convective boundary mixing, and ii) rotation. 

Convective boundary mixing (CBM) is a collective term that includes 
convective entrainment, convective penetration, and convective 
overshooting, as well as shear instabilities, and the generation 
internal gravity waves \citep{viallet2015,cristini2015}. Convective entrainment 
is the process by which mass (and hence chemicals) is (are) transported 
into a convective region due to turbulent motion at the interface 
of a convective region with a stably stratified region \citep{meakin2007}. 
Although convective entrainment is commonly implemented in 3D 
hydrodynamic simulations, it has not yet been widely implemented 
in 1D stellar evolution codes \citep{staritsin2013}.

Convective penetration and convective overshooting represent two consequences 
of the same process, and in the literature are often both referred to as
overshooting. This process refers to a convective element passing beyond 
a convective boundary as set by the Schwarzchild or Ledoux criterion due to 
its inertia, by an amount scaled in terms of the local pressure scale height. 
In convective penetration, the extended region adopts the adiabatic temperature 
gradient, which makes the extended region fully mixed and alters the thermal 
structure of the star, thus effectively enlarging the convective region. 
However, in convective overshooting, the extended region adopts the radiative 
temperature gradient, altering only the chemical structure of the equilibrium 
model. Due to limits in their implementation in 1-D stellar models, these 
phenomena produce the same effect on non-asteroseismic observables, to differing 
degrees \citep{godart2007}. A complete CBM profile would consist of entrainment, 
penetrative, and overshooting profiles stitched to the Schwarzschild boundary 
\citep{hirschi2014}, but this has yet to be consistently implemented in 1-D 
stellar models. Additional mechanisms that can contribute to CBM currently 
lack firm observational characterisation. Due to degeneracies in their implementation,
it can be more sensible to only consider a single mechanism that contributes to 
the overall CBM profile, such as convective overshooting as described above, 
and measure those stellar quantities altered by different amounts of overshooting. 

Rotationally induced mixing and its impact on stellar evolution is a 
heavily researched area in single stars \citep{abt2002,ekstrom2012,zhang2013}, 
binary stars \citep{torres2010,schneider2014,brott2011a,brott2011b,deMink2013}, 
pulsating stars \citep{vanReeth2016,vanReeth2018,ouazzani2017,christophe2018}, 
and stellar clusters \citep{evans2006,niederhofer2015,ahmed2017,bastian2017}. 
However, in 1-D stellar models, rotational mixing is often introduced 
as a diffusive mixing term, making its contribution degenerate with 
those CBM process previously described, as well as with any additional 
mixing term implemented. For a comprehensive overview of rotation in 
stellar astrophysics, we refer the reader to \cite{maeder2009}. 

Convective overshooting (or CBM in general) alters the chemical 
($\mu$)-gradient in the near-core region and increases the mass of the core. 
The $\mu$-gradient is important for massive stars that have a shrinking 
convective core as they evolve along the MS. The most important observational 
quantity varies from one modelling technique to another. For binaries and 
stellar clusters, the amount of CBM extends MS lifetimes, and hence alters 
the effective temperature, radius, and surface gravity of a star at a given 
age compared to when no CBM is present. For asteroseismic studies of pulsating stars, the $\mu$-gradient 
and core mass dictate the characteristics of the gravity (g)-mode cavity, and hence the 
observed pulsation periods. Thus, the important 
quantity is the core mass of a star along its MS, but this is not an input 
parameter for the computation of stellar models. Rather, it is an output 
parameter once the overshooting description, $D_{\rm ov}$, has been chosen. 
Therefore, estimation of $D_{\rm ov}$ (both the initial mixing 
value $D_0$ and the shape of the overshoot description along the radial 
coordinate) is necessary to derive the core mass, $\mu$-gradient and age of 
the star along the MS. 

While the investigation of the consequences of an enhanced core mass for stars
in a binary system extends back several decades \citep[see e.g.][for some early 
works]{zahn1977,roxburgh1978,maeder1987,andersen1990,zahn1991}, today it 
is largely seen as a necessity to find agreement between observed dynamic masses 
and evolutionary masses. This so-called mass discrepancy is encountered when
evolutionary tracks computed at the spectroscopic mass or dynamic mass obtained 
via eclipse modelling cannot reproduce spectroscopic temperatures or dynamic 
radii and surface gravities of the binary system \citep{herrero1992,tkachenko2014}. 
To match these observed quantities, a more massive star or a star of the same 
mass with  a more massive core is required. Furthermore, these quantities must 
be matched at the same age for both components in the system. This can be done 
by fitting individual tracks and enforcing the same age at each evaluation, 
or by fitting stellar isochrones. 

Several studies spanning stars of a considerable mass range have noted the need
for at least some amount of overshooting to reconcile the otherwise discrepant
dynamic and evolutionary masses
\citep{claret_gimenez1991,schroder1997,iwamoto1999,ribas2000,torres2010,tkachenko2014,claret2018}. 
Recently, \cite{claret2016,claret2017,claret2018} (hereafter  CT16, CT17, and CT18),
respectively) have explored the mass dependence of overshooting in a sample of
well detached, evolved, double-lined (SB2) eclipsing binaries (EBs). The authors
computed several tracks at the determined dynamic mass with varied overshooting 
(a step-overshooting prescription: $\alpha_{\rm ov}$ in CT16, and a diffusive exponential 
description: $f_{\rm ov}$ in CT17 and CT18) and $\alpha_{\rm MLT}$ 
values, then fit the tracks individually according to their respective observed 
quantities allowing for a 5~per~cent difference in age between the two components. 
Their results revealed an apparent mass dependence of overshooting from 1.2 to 2 
${\rm M_{\odot}}$ with no significant mass dependence from 2 to 4 ${\rm M_{\odot}}$. 
However, CT16, CT17, and CT18 did not take into account important degeneracies amongst 
the stellar parameters. Indeed, it is well known from gravity-mode asteroseismology 
that the core mass, via the core overshooting, is not only degenerate and correlated 
with the mass of the star, but also with metallicity and central hydrogen content 
\citep{moravveji2015,moravveji2016,schmid2016,buysschaert2018b}. Moreover, estimation of $D_{\rm ov}$ also requires one to take 
into account the depedencies of the choice of nuclear network, chemical mixture, opacity 
tables, atomic diffusion \citep[e.g.][]{aerts2018b}. Only a systematic approach taking into account these 
parameter degeneracies can lead to proper estimates of core masses and ages of stars. 
This is supported by the work of \cite{constantino2018} who show that the need for 
overshooting in evolutionary tracks is less obvious when considering the sensitivity 
of commonly used observables to varying amounts of overshooting. Such a 
systematic study has not yet been done for binaries, which is the topic of this paper.

The extended MS turn-off (eMSTO) refers to the spread in effective temperature and surface
gravity (or color and magnitude) observed in the turn-off point of open and globular clusters.
Historically, this has been investigated as being caused by several populations of stars with 
different rotation rates, where an entire cluster is assumed to have a singular amount of 
overshooting \citep{bastian2017}. Regarding the Large Magellanic Cloud, convective 
core overshooting, metallicity, extinction, distance, and age were estimated simultaneously 
from Hubble Space Telescope observations by \citet{rosenfield2017}. This led to $D_{\rm ov}$ 
values in line with canonical values, but with a proper (large!) uncertainty range. 
\citet{yang2017} have recently interpreted the eMSTO as being partly caused by stars with 
varying values of $D_{\rm ov}$. However, the authors did not systematically account for
degeneracies with other parameters connected with the choice of the input
physics in their stellar models. 

Gravity mode oscillations are excellent calibrators for near-core mixing
processes, such as core overshooting, because they propagate in the deep
stellar interior near the core, and are sensitive to the processes at
work there. For example, g modes have been used to estimate the near-core 
rotation rate in some 40 BAF-type stars, covering the range of very
slow rotation to half critical \citep{kurtz2014,saio2015,murphy2016,vanReeth2016,vanReeth2018}. 
This was achieved by exploiting the properties of g-mode period spacings which
were first discovered in MS Slowly Pulsating B (SPB) stars by the CoRoT mission \citep{degroote2010,papics2012} 
and have since been observed in {\it Kepler\/} space photometry of numerous SPB and $\gamma$ Dor stars 
\citep{vanReeth2015b,bedding2015,papics2017,ouazzani2017}. 

Period spacings of g modes are constructed by taking the difference between the
periods of modes with the same degree, $\ell$, and azimuthal order, $m$ \citep{aerts2010}. 
Gravity modes and hence g-mode period spacing patterns are sensitive to the 
mass of the core, which sets the scaling of the pattern for a given mode geometry 
($\ell,m$). Deviations from uniformity are produced by trapped g modes, 
whose frequencies are bumped due to the near-core $\mu$-gradient that results 
from internal mixing \citep{miglio2008}. Additionally, g modes are deflected by 
the Coriolis force according to their geometry, producing a negative slope in 
period spacing patterns of prograde and zonal g modes and a positive slope in the patterns 
of retrograde g modes \citep{miglio2008,bouabid2013}. This theoretical interpretation 
has been used to interpret period spacing patterns detected in space 
photometry in terms of near-core rotation, overshooting, and diffusive envelope mixing 
\citep{vanReeth2016,moravveji2015,moravveji2016,schmid2016,ouazzani2017,papics2017}. 
Recently, period-spacing patterns have been modelled to reveal the  shape and 
extent of core overshooting in a handful of SPB and $\gamma$ Dor stars
\citep{moravveji2015,moravveji2016,schmid2016,buysschaert2018b,szewczuk2018}. 
Additionally period spacing patterns allowed \citet{aerts2017} to place the 
detected near-core rotation rates into an evolutionary sequence, with the 
aim to remedy shortcomings in angular momentum transport inside stars.

Assuming the detection of at least one g-mode period spacing pattern, one can simultaneously 
estimate the near-core rotation rate and the asymptotic period spacing value $\Pi_0$, 
given by:
\begin{equation}\label{eq:pi0}
  \Pi_0 = 2\pi^2\left( \int \frac{N}{r} \rm{d}r \right)^{-1}.
\end{equation}
This quantity is sensitive to any phenomenon that alters the spatial 
distribution of the Brunt-V\"ais\"al\"a frequency $N$, defined as: 
\newpage
\begin{equation}\label{eq:bruntV}
     N^2\simeq \frac{g^2 \rho}{P} \left( 
     \nabla_{\rm ad} - \nabla + \nabla_{\mu}  \right)
\end{equation}
\citep{miglio2008}. Thus, processes affecting the cavity where $N$ 
is positive, as well as the density and the chemical gradient near 
the core, will change the evaluation of $\Pi_0$. As both a star's 
evolution and $\Pi_0$ are sensitive to core overshooting, the combined 
modelling of these provides the unique opportunity to impose constraints 
on core overshooting.

Despite their complementary nature and the extensive discussion of their synergies
in the literature \citep{clausen1996,decat2000,decat2004,aerts2004,miglio2005},
simultaneous binary and asteroseismic modelling efforts to investigate interior
mixing have rarely been achieved. 
In the case of solar-like oscillators, a handful of MS, subgiant and red giant
binaries, where both components exhibit oscillations, have been modelled 
\citep{miglio2005,appourchaux2014,appourchaux2015,metcalfe2015,white2017,bellinger2017,li2018,beck2018a}.
With the exception of \cite{beck2018a}, all of these studies modelled the systems individually and 
employed the assumption of equal age and initial chemical composition as an {\it a posteriori}
test. Red giant binaries with one pulsating component are more common \citep{beck2014,themesl2018}. \cite{beck2014} 
simultaneously investigated the binary and asteroseismic signals in KIC 5006817 to study the
angular momentum and dynamical evolution of that unresolved binary system.
The most recent example of combined binary and asteroseismic modelling was carried 
out for the almost twin $\delta$ Sct/$\gamma$ Dor binary system KIC\,10080943 by \cite{schmid2016}. In this study,
the authors were able to identify multiple g-mode period spacing patterns 
corresponding to the individual components of the binary and carry out asteroseismic 
modelling. Furthermore, \cite{schmid2016} enforced equal age in 
the modelling procedure, rather than as an {\it a posteriori} test

This paper is part of a larger extensive study of parameter estimation and stellar 
model selection, which properly takes into account correlations and degeneracies, 
of single and binary stars that pulsate in g modes. Here, we take the  
first steps in developing a methodology which integrates binarity and gravity-mode 
asteroseismology and show that the incorporation of binary information in asteroseismic 
modelling changes the solution space and refines the model selection process. 
We provide a framework for such simultaneous 
asteroseismic and binary modelling of g-mode pulsators in binaries using isochrones 
based on $\Pi_0$ to quantify the extent of near-core mixing attributed to convective 
core overshooting, as well as the mass and size of the convective core. We do this in 
the simplest case where the two stars are assumed to have the same initial 
metallicity, $Z=0.014$ and adopt one prescription for the core overshooting, keeping 
in mind the known degeneracies among those quantities and stellar mass and age. Later 
studies will consider more complex isochrone construction where the full parameter 
space covering the most important phenomena needed to interpret the g-mode frequencies as 
discussed in \citet{aerts2018b} will be taken into account. In Section 2 we discuss 
the stellar models we computed and in Section 3 we discuss the construction of 
isochrones and outline our modelling methodology. In Sections 4 and 5 we present 
applications to three binary systems observed by {\it Kepler} and discuss the 
implications of our integrated modelling 
approach compared to the case where we treat the binary components as single stars. 
Finally in Section 6 we summarize the strategy for future work. Our isochrones 
are made available electronically for use by the community.

\section{Stellar models}
To evaluate the binary and asteroseismic properties of our target stars, we
construct a grid of stellar evolutionary models, using the open source MESA
software \cite[r10108;][]{mesaI,mesaII,mesaIII,mesaIV}. The goal of this grid is
to obtain reliable estimates of fundamental stellar parameters, such as the
initial mass, the core hydrogen content, the core mass, and the extent of the
overshooting region. To adequately assess these quantities, we must consider
several model parameters and input physics choices, which are discussed below. 
To inspect if there is a benefit in joint binary asteroseismic modelling, we 
establish the necessary lowest possible 
dimensionality. We hence fix the initial hydrogen and helium fractions ($X,Y$), 
as well as the metallicity $Z$.  We do not treat binary evolution in our grid of
equilibrium models, but instead assume each star has undergone secular single
star evolution. After our exploration of the benefit of binary asteroseismic
modelling, more general model grids for isochrone construction will be
considered, as well as applications to g-mode pulsators in eclipsing binaries
and clusters.
\begin{table}
	\centering
	\caption{The range and step for all parameters varied in the
          stellar model grid.}
	\label{tab:grid_pars}
	\begin{tabular}{lcccr}
		\hline
		 & Unit & Lower & Upper & Step\\
		\hline
		$\mathrm{M_{ini}}$ & $\mathrm{M_{\odot}}$ & 1.2 & 10 & 0.1\\
         $f_{\rm ov}$ & & 0.005 & 0.04 & 0.005 \\
		\hline
	\end{tabular}
\end{table}
\subsection{Convection and overshooting}
Compared to evolutionary time scales of stars, convective phenomena are
instantaneous. This difference in time scales allows for a simplification 
in the implementation of convection in evolution codes. In terms of element 
transport, it is appropriate to consider the mixing due to convection as 
instantaneous mixing over a scale distance in regions that satisfy a convection 
criterion. One of the most widely used implementations of convection in 1-D 
is mixing length theory (MLT) \citep{bohmVitense1958} and its variations. 
MESA employs several variations of MLT to model convection \citep{mesaIV}.
Through this formalism, convection is considered to be very efficient mixing 
with the effective mixing distance being scaled by the parameter $\alpha_{\mathrm{MLT}}$. 
Recent asteroseismic modelling efforts in intermediate-mass stars have fixed 
$\alpha_{\rm MLT}=2.0$ \citep{moravveji2016}. Initially, we consider this as 
the center point of our grid step in $\alpha_{\rm MLT}$. 

A major limitation of MLT is its inability to predict the behaviour 
of convective fluid elements at the boundary of a convective region, defined 
by the Schwarzschild criterion \citep[e.g.][]{kippenhahn2012}. Due to their 
inertia, the convective elements cannot abruptly stop when they move from a 
convective region to a radiative region. Thus, they ``overshoot'' the boundary, 
causing mixing in that transition zone.

Implemented as a means to remedy the theoretical shortcomings of most 1-D
convective descriptions, convective overshooting has received much 
attention in the past, but remains poorly calibrated by observations. 
Overshooting generally follows one of two major descriptions: 1) step 
overshooting:
\begin{equation}\label{eq:step_ov}
D_{\rm ov}=\alpha_{\rm ov} H_{\rm p},
\end{equation}
where $H_{\rm p}$ is the local pressure scale height, and $\alpha_{\rm ov}$ 
is the extent by which the overshooting region extends, 
and 2) exponential overshooting:
\begin{equation}\label{eq:exp_ov}
    D_{\rm ov}=D_0 \exp{\left(\frac{-2(r-r_0)}{f_{\rm ov}H_{\rm p}}\right)}.
\end{equation}
Here, $D_0$ is the diffusive coefficient at the radial coordinate $r_0$ where 
the exponential profile begins. It is important to note that the overshoot 
region uses the radiative temperature gradient, meaning that only chemical 
mixing occurs without altering the thermal structure of the model. In our models, 
we do not consider the mass present in the overshooting region in the calculation 
of the convective core mass, ${\rm M_{cc}}$. However, overshooting still changes 
${\rm M_{cc}}$ by supplying hydrogen from a stably stratified region into the
instantaneously mixed convective zone, which extends the MS lifetime and alters
the resulting helium core mass at the terminal age MS (TAMS). Additionally, the 
chemical mixing alters $\nabla_{\mu}$ just outside of the core according to the 
chosen prescription.

Recent asteroseismic modelling by \cite{moravveji2015,moravveji2016} has shown that
a diffusive exponential overshooting better reproduces observed period-spacing
patterns of B-type stars of $\sim3 {\rm M_{\odot}}$ compared to a diffusive step
overshooting implementation -- see also \citet{pedersen2018} for the capacity of
g modes to distinguish these two from mode trapping. Here, we use diffusive 
exponential overshooting in our grid, taking the values shown in Table 
\ref{tab:grid_pars}. 
While both diffusive overshooting and convective penetration may be simultaneously 
active in stars, we consider this configuration in future work. Here,
we only consider convective overshooting in our models 
and use it as a proxy for the total amount of CBM with the aim to estimate the 
convective core mass and size of stars.

Since r10000, MESA has a scheme to robustly determine the convective core
boundary, which ensures a continuous temperature gradient from the convective
core to the radiative envelope \citep{mesaIV}. We adopt this predictive mixing
scheme in this work. Given that our grid and isochrones are computed for both receding
and growing convective cores, we select the Ledoux criterion for convection to
account for the chemical gradient present in the case of a shrinking convective
core.

\subsection{MESA model physics and inlist}
Varying the metallicity within a grid of evolutionary tracks introduces the aforementioned 
degeneracy to modelling: lower metallicity tracks shift to higher temperatures 
and thus effectively mimic evolutionary tracks produced with higher masses. 
This causes a near perfect degeneracy between initial mass and metallicity in 
evolutionary modelling. As such, we fix the metallicity, selected to be the 
cosmic B-star metallicity $Z=0.014$ \citep{pryzbilla2008,nieva2012}. We fix 
the initial helium fraction to $Y=0.276$ taken from the cosmic B-star solution 
and solve for initial hydrogen fraction as $X=1-0.276-0.014$. In doing so we 
fix the mean molecular weight contribution from helium, which would otherwise 
contribute to the mass-metallicity degeneracy. We enforce that both stars in 
a system being evaluated have the same initial metallicity, which is a reasonable 
assumption as the two components of a given system collapse from the same proto-stellar 
cloud. We choose a mass range that spans 1.2 to 10 ${\rm M_{\odot}}$, 
which covers the mass range where g-mode pulsations are theoretically expected 
for stars with a convective core. Additionally, we vary the extent of overshooting, 
$f_{\rm ov}$, to cover the entire range encountered so far from asteroseismic modelling. 

Due to the limitations of its parametrized implementation, exponential diffusive 
overshooting alone was not sufficient to accurately model the observed period 
spacing patterns of KIC\,10526294 and KIC\,7760680 \citep{moravveji2015,moravveji2016}. 
In both cases, to better reproduce the observations the authors needed to include 
an additional diffusive mixing term, $D_{\rm mix}$, which alters the chemical 
stratification in the envelope of the star where the period spacing pattern is most 
sensitive. As this parameter alters the core mass, we are interested in varying this 
quantity in our grid, but restrict it to less than $10^3 \, {\rm cm^2\,s^{-1}}$ as \cite{pedersen2018} 
{\rm have} shown that values larger than this wash away the chemical gradient outside of the 
core and destroy mode trapping which is observed in period-spacing patterns. The contribution 
of any additional mixing term is implemented as a diffusive coefficient, making it 
degenerate with $D_{\rm mix}$. Thus, we consider $D_{\rm mix}$ to be a catch-all extra 
mixing term to be calibrated.

The impact of rotation on seismic modelling cannot be ignored at the level of pulsation 
mode modelling. Rotation is introduced at the level of the pulsation eigenmode calculations 
in 3-D. Recently, \cite{aerts2018b} have shown that eigenmodes computed with and without the 
Coriolis force (at 10~per~cent critical rotation rate) have differences larger 
than the observational frequency uncertainty provided by the 4-yr time base of the nominal
{\it Kepler} Space Telescope 
\citep{borucki2010}. Additionally, MESA implements rotationally induced chemical mixing 
as a diffusive term, which is degenerate with the extra diffusive mixing term $D_{\rm mix}$ 
that we already include. As stated previously, since we use $\Pi_0$ as our seismic diagnostic
and do not attempt to model individual mode frequencies, we choose to compute non-rotating 
hydrostatic-equilibrium stellar evolution models. 

As discussed by \cite{aerts2018b}, the choice of 1-D equilibrium model physics such as opacity, 
chemical mixture, amongst others, alters seismic modelling. All of our MESA tracks are computed 
from the Hayashi track, with the fully extended CNO nuclear network option 
``pp\_cno\_extras\_o18\_ne22.net", a simple photosphere, and do not include atomic diffusion. We 
assume a uniform initial chemical composition, fix the element fraction for the metals  
to those of \cite{asplund+2009}, and use the MESA defaults for the equation-of-state and opacities. 
We do not include mass loss and only calculate non-rotating models as discussed previously. 
Additionally, we use the \cite{cox_giuli_1968} implementation of the mixing length theory for 
convection and employ the Ledoux criterion for convective stability combined with the new predictive 
mixing scheme in MESA. Our base inlist is posted in the MESA Marketplace:
\url{http://cococubed.asu.edu/mesa_market/inlists.html}.

\subsection{Seismic diagnostic}
Our seismic diagnostic is applied in the form of the asymptotic period spacing $\Pi_0$ 
(Eq. \ref{eq:pi0}), which can be computed directly from the MESA models. As previously 
discussed, $\Pi_0$ is sensitive to any process that has an impact on the Brunt-V\"ais\"al\"a 
frequency. As such, we are interested in probing convective-core overshooting, $\alpha_{\rm MLT}$, 
and additional mixing in our grid. 

The g modes are sensitive to the mass and radius of the convective core, which depend on 
the age and mass of the star, and overshooting, which directly alters the core mass. The traditionally defined convective 
core radius is not altered by the diffusive exponential overshooting description since the 
thermal structure of the star is not altered. One can re-define the radius, and hence mass, 
of the core to lie at the boundary of the overshooting region, as was done by CT16 and CT17. 
However we define the core as at the position of the Schwarzschild boundary. Hence, the core 
mass will still account for the influx of mass due to near-core mixing as the star evolves. 
The sensitivity of $\Pi_0$ to $\alpha_{\rm MLT}$ and $D_{\rm mix}$ is less obvious. 
Table \ref{tab:dP_sensitivity} lists the difference in $\Pi_0$ for different parameter 
combinations at different combinations of masses and $X_c$ compared to a baseline. This table 
clearly demonstrates that constraining $\alpha_{\rm MLT}$ and $D_{\rm mix}$ only becomes 
possible by modelling the observed trapping properties ofindividual modes. Following this, 
we only investigate $f_{\rm ov}$ in our calculations, leaving $\alpha_{\rm MLT}$ 
and $D_{\rm mix}$ as nuisance parameters which can be fixed or marginalised over.

\begin{table*}
	\centering
	\caption{Differences in $\Pi_0$ from one parameter combination to a baseline 
	         at fixed masses, and $X_C$ combinations. Baseline denoted by *.}
	\label{tab:dP_sensitivity}
	\begin{tabular}{lcccccc}
		\hline
		${\rm M\, [M_{\odot}]}$ & $X_C$ & ${\rm Z_{ini}}$ & ${\alpha_{\rm MLT}}$ & $f_{\rm ov}$ & $\log D_{\rm mix} \, {\rm[cm^2 s^{-1}]}$ & $\delta\Pi_{0}\,{\rm[s]}$\\
		\hline
        1.5 & 0.70 & 0.0140 & 2.0 & 0.0200 & 1.0 & 0*  \\
        1.5 & 0.70 & 0.0140 & 2.0 & 0.0250 & 1.0 & 20 \\
        1.5 & 0.70 & 0.0140 & 1.8 & 0.0200 & 1.0 & 2  \\
        1.5 & 0.70 & 0.0140 & 2.0 & 0.0200 & 1.5 & 11 \\
        4.5 & 0.70 & 0.0140 & 2.0 & 0.0200 & 1.0 & 0*  \\
        4.5 & 0.70 & 0.0140 & 2.0 & 0.0250 & 1.0 & 2  \\
        4.5 & 0.70 & 0.0140 & 1.8 & 0.0200 & 1.0 & 1  \\
        4.5 & 0.70 & 0.0140 & 2.0 & 0.0200 & 1.5 & 2  \\
        9.0 & 0.70 & 0.0140 & 2.0 & 0.0200 & 1.0 & 0*  \\
        9.0 & 0.70 & 0.0140 & 2.0 & 0.0250 & 1.0 & 5  \\
        9.0 & 0.70 & 0.0140 & 1.8 & 0.0200 & 1.0 & 0  \\
        9.0 & 0.70 & 0.0140 & 2.0 & 0.0200 & 1.5 & 3  \\
    	\hline
        1.5 & 0.30 & 0.0140 & 2.0 & 0.0200 & 1.0 & 0*   \\
        1.5 & 0.30 & 0.0140 & 2.0 & 0.0250 & 1.0 & 51  \\
        1.5 & 0.30 & 0.0140 & 1.8 & 0.0200 & 1.0 & 30  \\
        1.5 & 0.30 & 0.0140 & 2.0 & 0.0200 & 1.5 & 15  \\
        4.5 & 0.30 & 0.0140 & 2.0 & 0.0200 & 1.0 & 0*   \\
        4.5 & 0.30 & 0.0140 & 2.0 & 0.0250 & 1.0 & 127 \\
        4.5 & 0.30 & 0.0140 & 1.8 & 0.0200 & 1.0 & 10  \\
        4.5 & 0.30 & 0.0140 & 2.0 & 0.0200 & 1.5 & 59  \\
        9.0 & 0.30 & 0.0140 & 2.0 & 0.0200 & 1.0 & 0*   \\
        9.0 & 0.30 & 0.0140 & 2.0 & 0.0250 & 1.0 & 202 \\
        9.0 & 0.30 & 0.0140 & 1.8 & 0.0200 & 1.0 & 4   \\
        9.0 & 0.30 & 0.0140 & 2.0 & 0.0200 & 1.5 & 6   \\
    	\hline
        1.5 & 0.05 & 0.0140 & 2.0 & 0.0200 & 1.0 & 0*   \\
        1.5 & 0.05 & 0.0140 & 2.0 & 0.0250 & 1.0 & 54  \\
        1.5 & 0.05 & 0.0140 & 1.8 & 0.0200 & 1.0 & 1   \\
        1.5 & 0.05 & 0.0140 & 2.0 & 0.0200 & 1.5 & 14  \\
        4.5 & 0.05 & 0.0140 & 2.0 & 0.0200 & 1.0 & 0*   \\
        4.5 & 0.05 & 0.0140 & 2.0 & 0.0250 & 1.0 & 166 \\
        4.5 & 0.05 & 0.0140 & 1.8 & 0.0200 & 1.0 & 6   \\
        4.5 & 0.05 & 0.0140 & 2.0 & 0.0200 & 1.5 & 70  \\
        9.0 & 0.05 & 0.0140 & 2.0 & 0.0200 & 1.0 & 0*   \\
        9.0 & 0.05 & 0.0140 & 2.0 & 0.0250 & 1.0 & 296 \\
        9.0 & 0.05 & 0.0140 & 1.8 & 0.0200 & 1.0 & 2   \\
        9.0 & 0.05 & 0.0140 & 2.0 & 0.0200 & 1.5 & 11  \\
		\hline
	\end{tabular}
\end{table*}

\section{Methods}
In this section we outline our methodology for isochrone construction 
and selection of valid models for seismic evaluation according to the 
equal age and initial chemical composition constraints and mass ratio 
enforced by binarity, while allowing for differing model parameters 
for the primary and secondary components. 

\subsection{Isochrone construction}
\begin{figure}
    \includegraphics[width=\columnwidth]{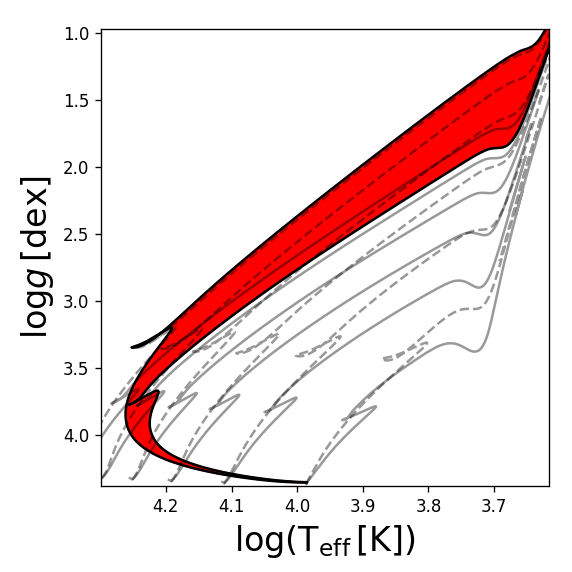}
    \caption{Example evolutionary tracks (grey solid and dashed lines), isochrones (black solid lines),
     and isocloud (red x-markers). All evolutionary tracks are computed with $Z_{\rm ini}=0.014$ and 
     $\alpha_{\rm MLT}=2.0$. Evolutionary tracks range from 2 to 7 ${\rm M_{\odot}}$ with 
     $f_{\rm ov}=0.005$ (solid grey lines) and $f_{\rm ov}=0.040$ (dashed grey lines).}
    \label{fig:example_ischrones}
\end{figure}

After constructing our grid with parameters as defined in Table 
\ref{tab:grid_pars}, we build isochrones according to \cite{dotter2016}. 
This methodology involves two steps. In the first step, in every 
evolutionary track we identify the main phases of evolution from Pre-MS 
(PMS) to zero-age MS (ZAMS), ZAMS to middle-age MS (MAMS), MAMS to TAMS, 
and TAMS to the onset of core helium burning (RGBhb). From there, we 
identify $n$ equidistantly spaced steps between any two main evolutionary 
phases according to a weighting function accounting for the change in 
$\log T_{\rm eff}$, $\log g$, $\log T_c$, and other quantities. The 
tracks are then interpolated to these points to create equivalent 
evolutionary phase (EEP) tracks. In the second step, we simply loop over 
a given EEP point in all tracks to construct a monotonic mass-age 
relationship. From this relationship, we then build our isochrones by 
interpolating in all quantities that we are interested in, using mass as 
the independent variable. For an elaborate discussion and detailed 
testing of this method, we refer the reader to \cite{dotter2016} and 
\cite{choi2016}.

Traditionally, an isochrone at a given age $\tau$ is constructed from 
evolutionary tracks that were all computed with the same input physics 
and equal free parameters, i.e. with the same $\Theta_{\rm iso} = 
\left(\mathrm{Z_{\rm ini}}, \alpha_{\rm MLT}, f_{\rm ov}, D_{\rm mix}, 
\tau \right)$. However, changing the core overshooting in a model alters 
the evolutionary tracks, and hence, the isochrones constructed from these 
tracks. Figure \ref{fig:example_ischrones} highlights the difference in 
isochrones (solid black lines) constructed from tracks with  $f_{\rm ov}=0.005$ 
(right-most isochrone) and $f_{\rm ov}=0.040$ (left-most isochrone), constructed 
from the solid grey tracks and dashed grey tracks, respectively. These two 
limiting values of $f_{\rm ov}$ result from asteroseismology of various pulsators
\citep{briquet2007,moravveji2015,moravveji2016,buysschaert2018b}. If one were 
to use isochrones constructed with only one set of parameter combination, 
they would artificially restrict the solution to a singular region of the 
parameter space, or potentially drive their solution space to an unrealistic 
range. Instead, we introduce the concept of an isochrone-cloud (hereafter isocloud), 
i.e. the collection of all isochrones 
(created from all combinations of ${\rm Z_{ini}}, \alpha_{\rm MLT}, f_{\rm ov}, 
D_{\rm mix}$) calculated at a given age $\tau$, given by: $\Theta_{\rm cloud} 
= \left({\rm Z_{ini}}, \alpha_{{\rm MLT-}i}, f_{{\rm ov-}j}, D_{\rm{mix-}k}, 
\tau \right)$, where $i$, $j$, and $k$, denote the different possible values 
of $\alpha_{\rm MLT}, f_{\rm ov}$, and $D_{\rm mix}$ respectively. An example 
isocloud for fixed $\alpha_{\rm MLT}$, and $D_{\rm mix}$ at a given age 
can be seen in Fig. \ref{fig:example_ischrones} as the 
area spanned in red. We note that this region falls between the two 
extreme cases for the traditional isochrones. 
We provide our isochrone-clouds to the community on VizieR\footnote{http://vizier.u-strasbg.fr/viz-bin/VizieR}.

\subsection{Forward modelling scheme}
\cite{aerts2018b} proposed a scheme for the forward seismic modelling 
of g-mode pulsators with a convective core that accounts for the 
degeneracies produced by combinations of varied free parameters in the 
modelling process. This work has recently been employed by Mombarg et al. 
({\it in prep.}) who use ${\rm T_{eff}}$, $\log g$, and $\Pi_0$ to estimate 
the age, mass, core mass, and core overshooting of $\gamma$ Dor stars after 
estimation of the near-core rotation rate by \cite{vanReeth2016}. 

Here, we adopt the same framework to estimate stellar parameters, but extend it to 
include binary information. This binary information comes in the form of equal age and initial chemical 
composition, the effective temperature and surface gravity of the secondary, 
as well as information on the component masses and radii. In the case of an EB
or heartbeat star \citep{welsh2011,thompson2012}, this comes in the form of direct 
estimates of dynamical masses and radii from binary modelling. However, in the case of a
SB2, binary modelling only yields estimates of the mass and radii ratios to be applied to our
modelling scheme. We thus assume that the near-core rotation
rate has been deduced from the data, allowing us to derive $\Pi_0$ from the
observed period spacing pattern.

Following the notation of \cite{aerts2018b}, our fixed model physics is
contained in the vector $\psi$, where we write a single model as 
$\mathcal{M}_j\left( \theta_j, \psi \right)$ with $\theta_j$ defined as:
\begin{equation}
    \label{eq:theta_star}
    \theta_j=\left( {\rm Age}, {\rm M}, f_{{\rm ov}} \right),
\end{equation}
being the vector of the $j$-th combination of parameters listed in Table 
\ref{tab:grid_pars}. Each grid point $\mathcal{M}_j\left( \theta_j, 
\psi \right)$ with $j=1...{\rm N}$, with N being the total number of 
grid points, has corresponding values for ${T_{\rm eff}}$, $\log g$ 
and $\Pi_0$, written as:
\begin{equation}
{\bf Y}_j=\left( {\rm T_{eff}}, \log g, \Pi_0 \right),
\end{equation}
to be compared against the observed values of ${\rm T_{eff}}$, $\log g$ 
and $\Pi_0$ contained in the vector ${\bf Y}^*$ with associated 
uncertainties $ \varepsilon^*$. 

The extension to the binary case involves a change of basis to the 
isoclouds, which have age as an explicit model parameter and mass as 
an implicit variable. In this case, $\psi$ remains the same, but 
$\theta_j$ for a given point amongst any isocloud is now written as:
\begin{equation}
    \label{eq:theta_iso}
    \theta_j=\left( {\rm Age}, {\rm M}_i, f_{{\rm ov},i} \right)~;~i=1,2.
\end{equation}
Thus, a corresponding grid-point $\mathcal{M}_j\left( \theta_j, \psi \right)$ would
lead to the predicted vector:
\begin{equation}
    {\bf Y}=\left( {\rm T}_{{\rm eff},i}, \Pi_{0,i}, {\rm M}_i, {\rm R}_i \right)~;~i=1,2, 
\end{equation}
or 
\begin{equation}
	\label{eq:y_sb2}
    {\bf Y}=\left( {\rm T}_{{\rm eff},i}, \Pi_{0,i}, q, \mathcal{R}\right)~;~i=1,2, 
\end{equation}
depending on whether the system is an EB/heartbeat star, or SB2, respectively. 
In Eqn.~\ref{eq:y_sb2} $q={\rm \frac{M_2}{M_1}}$ is the mass ratio and $\mathcal{R}=
{\rm \frac{R_2}{R_1}}$ is the radii ratio. These are compared against the 
observations ${\bf Y}^*$ with errors $\varepsilon^*$ for either the EB/heartbeat star or SB2 
case. Our grid is constructed as the combination of every point in an isocloud 
with every other point of that isocloud at a given age, if at least one point 
falls within $3-\sigma$ of the observed values of ${\rm T_{eff}}$ and $\log g$ 
for the primary and secondary. This results in excess of a few million combinations. 
This configuration allows us to enforce that, while both components have the same 
age and initial chemical composition, they can have different amounts of $f_{\rm ov}$, 
as this reflects results from binarity (CT18) and asteroseismology 
\citep{aerts2015,moravveji2015,moravveji2016}. 
We adopt the Mahalanobis distance (MD) for our merit function, as described in  
\cite{aerts2018b}. The MD needs no modification for the current application.
The MD is calculated for every grid point, resulting in a distribution that approaches a 
$\chi^2$-distribution in the limiting case of approximately normally distributed parameters, whether they
are correlated or not. Based on this, we consider the 50th percentile as a good approximation to derive
confidence intervals for the input model parameters. This corresponds to $1-\sigma$ in the case of a 
normally distributed parameter distribution.
In the absence of prior information and posterior distributions, a Bayesian approach 
essentially reduces to a Maximum Likelihood Estimator (MLE).

\subsection{Hare-and-hound}

\begin{figure*}
    \centering
    \subfloat[\label{fig:hare_single-MD}]{\includegraphics[width=0.45\textwidth]{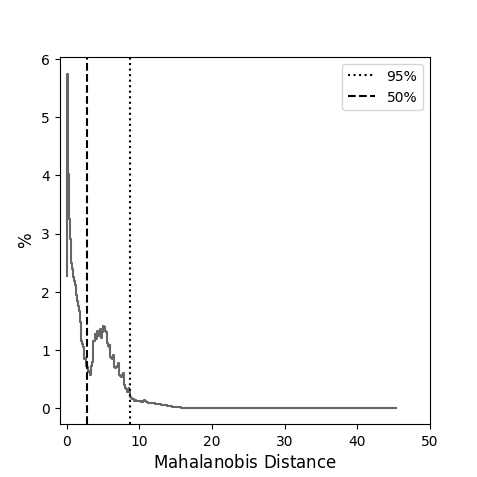}}
    ~
    \subfloat[\label{fig:hare_single-corrs}]{\includegraphics[width=0.45\textwidth]{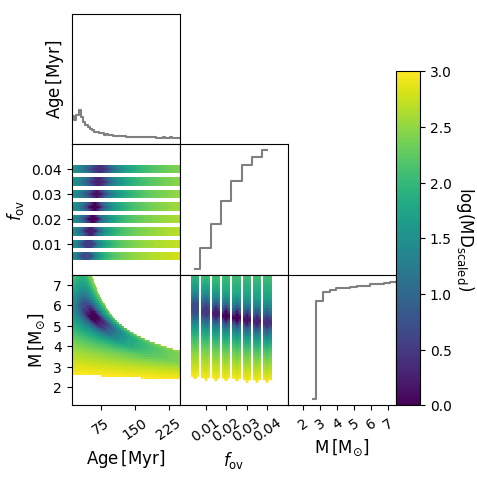}}
    \caption{\label{fig:hare_single-all}Mahalanobis distance distribution (left) and 50th inter-quartile ranges (right)
             for the primary of the hare-and-hound exercise using a single-star evaluation. All points in the correlation 
             plots are color coded according their rescaled Mahalanobis distance. Diagonal plots are binned parameter distributions. }
\end{figure*}

\begin{figure*}
    \centering
    \subfloat[\label{fig:hare_sb2-MD}]{\includegraphics[width=0.45\textwidth]{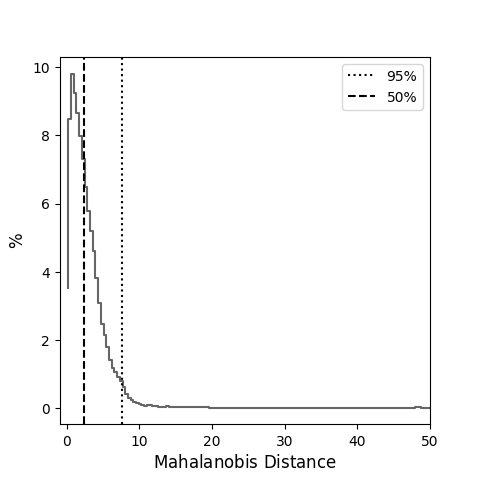}}
    ~
    \subfloat[\label{fig:hare_sb2-corrs-primary}]{\includegraphics[width=0.45\textwidth]{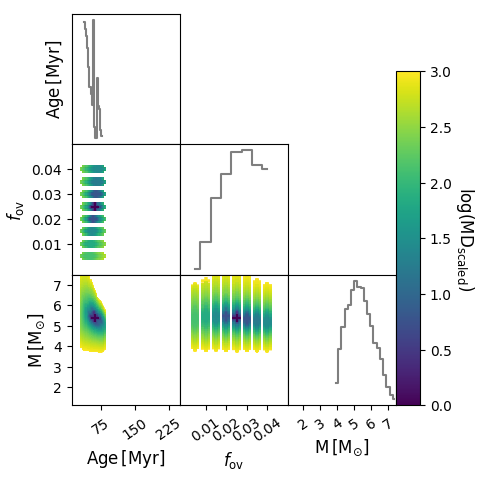}}
    \caption{\label{fig:hare_sb2-all} Same as Fig. \ref{fig:hare_single-all} but for the SB2 evaluation of the Hare-and-hound exercise. }
\end{figure*}
To test the methodology, we perform a hare-and-hound exercise, where the hare was computed from 
MESA using the input parameters listed in the top portion of Table \ref{tab:hare_pars}. The 
system is evaluated at an age of 59.7 Myr, resulting in the parameters seen in the bottom half 
of Table \ref{tab:hare_pars}. To simulate realistic observational errors for an SB2, we assume 
symmetric uncertainties of 200 K on ${\rm T_{eff}}$ for both stars, 300 s on $\Pi_0$, 0.01 on the mass 
ratio, and 0.05 on the radii ratio. 

The results of treating the hare as a single star and as the primary of an SB2 system can be 
seen in Figs. \ref{fig:hare_single-all} and \ref{fig:hare_sb2-all}. Fig. \ref{fig:hare_single-MD} 
shows the MD evaluations for each grid point. The vertical dotted line represents the 95th 
percentile and the dashed line represents the 50th percentile. Fig. \ref{fig:hare_single-corrs} 
shows the correlation structure for the components of the $\theta$ vector for all grid 
points with an MD below the 50th percentile cutoff. The binned distributions along the 
diagonal are projections of each component in $\theta$ onto one dimension. The colour 
represents the MD where all distances have been re-scaled between 1-1000 for ease of 
comparison across systems and cases. Since the MD is an MLE point estimator, we report 
the grid point with the lowest MD as the best model, where the MLE lower and upper bounds 
are taken as the minimum and maximum value of the parameter space within the 50th percentile 
(inter-quantile) cutoff, as listed in Table \ref{tab:hare_soln}. We are interested in other 
astrophysical quantities such as $\rm M_{cc}$, the radius ${\rm R}$, core hydrogen content 
$X_{\rm c}$, and radial location of the overshoot region ${\rm R_{ov}}$. However, since these 
quantities are not model input parameters in $\theta$, but rather are output of a given model, 
we list their values corresponding to the best model without corresponding MLE bounds.

In this example we can see that the binary evaluation greatly reduces the parameter space 
compared to the single-star evaluation. Furthermore, we can see that in both cases, the 
best models agree well with the input. 

\begin{table}
	\centering
	\caption{Input and target parameters for the hare-and-hound exercise.}
	\label{tab:hare_pars}
	\begin{tabular}{lcc} 
		\hline
		Parameter & Primary & Secondary \\
		\hline
		$\mathrm{M_{ini}\,[M_{\odot}]}$ & 5.40 & 3.83 \\
		$\mathrm{Z_{ini}}$ & 0.0135 & 0.0135 \\
		$\alpha_{\mathrm{MLT}}$ & 2.10 & 1.95 \\
		$f_{\rm ov}$ & 0.025 & 0.010 \\
		$D_{\mathrm{mix}}\,\mathrm{[cm^2\,s^{-1}]}$ & 20 & 15 \\
       \hline
		$\mathrm{T_{eff}\,[K]}$ & 16700 & 14450 \\
		$\log g\,\mathrm{[dex]}$ & 3.91 & 4.19 \\
		$\mathrm{R\,[R_{\odot}]}$ & 4.274 & 2.595 \\
		$\Omega_{\mathrm{rot}}\,\mathrm{[d^{-1}]}$ & 0.54 & -\\
		$\mathrm{Age\,[Myr]}$ & 59.7 & 59.7 \\
		$\mathrm{X_c}$ & 0.342 & 0.558 \\
		$\mathrm{\Delta\Pi\,[d^{-1}]}$ & 0.129 & 0.103 \\
		$\mathrm{M_{core}\,[M_{\odot}]}$ & 0.97 & 0.76 \\
       \hline
	\end{tabular}
\end{table}
\begin{table}
    \tabcolsep=2pt
	\centering
	\caption{MLE and $1-\sigma$ errors of the parameters for the Hare-and-hound exercise.}
	\label{tab:hare_soln}
	\begin{tabular}{lccc}
		\hline
		 & Parameter & Primary & Secondary \\
		\hline
        \multirow{5}{*}{Single} & $\mathrm{Age\,[Myr}]$ & \multicolumn{2}{c}{61  (0,  1593)} \\
		& $f_{\rm ov}$ &  0.025  (0.005,  0.04) & -- \\
		& $\mathrm{M\,[M_{\odot}]}$ & 5.4  (2.2,  9.9) & -- \\
		& $\mathrm{R\,[R_{\odot}]}$ & 4.31             & -- \\
		& $\mathrm{M_{cc}\,[M_{\odot}]}$ & 0.97        & -- \\
		& $\mathrm{R_{cc}\,[R_{\odot}]}$ & 0.50        & -- \\
		& $\mathrm{X_{c}}$               & 0.34        & -- \\
		\hline
       \multirow{5}{*}{SB2} & $\mathrm{Age\,[Myr}]$ & \multicolumn{2}{c}{63~(38,~78)} \\
		&                     $f_{\rm ov}$                    & 0.025~(0.005,~0.04) & 0.020~(0.005,~0.04) \\
		&                     $\mathrm{M\,[M_{\odot}]}$       & 5.37~(3.86,~7.48)   & 3.80~(2.58,~5.91) \\
		&                     $\mathrm{R\,[R_{\odot}]}$       & 4.33                & 2.57                \\
		&                     $\mathrm{M_{cc}\,[M_{\odot}]}$  & 0.96 & 0.78 \\
		&                     $\mathrm{R_{cc}\,[R_{\odot}]}$  & 0.49 & 0.39 \\
		&                     $\mathrm{X_{c}}$                & 0.34 & 0.58 \\
		\hline
	\end{tabular}
\end{table}
\section{{\it Kepler} Sample}
\begin{figure}
	\includegraphics[width=\columnwidth]{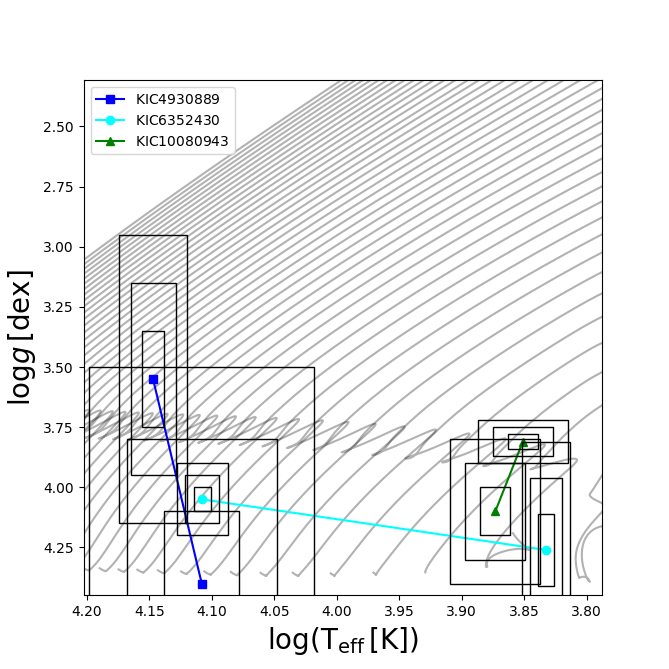}
    \caption{ Spectroscopic parameters for target binaries plotted over evolutionary 
    tracks. Tracks computed with $Z_{\rm ini}=0.014,\, \alpha_{\rm MLT}=2.0,\, f_{\rm ov}=0.015,\, D_{\rm mix}=10 \, {\rm cm^2\,s^{-1}}$.}
    \label{fig:sample_kiel}
\end{figure}

We apply our methodology to three g-mode pulsating stars observed with \textit{Kepler}. 
All targets are binary systems with at least one g-mode pulsating component with an 
estimate of $\Pi_0$, as measured according to \cite{vanReeth2016}, which have been previously 
studied in the literature  \citep{papics2013,schmid2015,schmid2016,papics2017}. The 
relevant spectroscopic and binary parameters are listed in Table \ref{tab:kepler_pars}. 
The three systems are plotted in a Kiel diagram in Fig. \ref{fig:sample_kiel}.

\begin{table*}
    \centering
    \caption{Measured spectroscopic, binary, and asteroseismic parameters for \textit{Kepler} targets.}
    \label{tab:kepler_pars}
    \begin{tabular}{lcccccc}
        \hline
        Parameter & \multicolumn{2}{c}{KIC 4930889} & \multicolumn{2}{c}{KIC 6352430} & \multicolumn{2}{c}{KIC 10080943}\\ 
        \hline
        $\mathrm{T_{eff} \, [K]}$           & $14\,020\pm280$ & $12\,820\pm900$    & $12\,810\pm200$ & $6805\pm100$       & $7150\pm250$ & $7640\pm240$ \\
        $\log g \, \mathrm{[dex]}$          & $3.55\pm0.24$   & $4.38\pm0.10$      & $4.05\pm0.05$   & $4.26\pm0.15$      & $3.81\pm0.03$ & $4.10\pm0.10$ \\
        ${\rm\Pi_0 \, [s]}$                 & $8712\pm320$    & --                 & $6944\pm900$    & --                 & $3984\pm38$ & $4108\pm51$ \\
        $q\, \mathrm{[\frac{M_2}{M_1}]}$    & \multicolumn{2}{c}{$0.77\pm0.09$}    &\multicolumn{2}{c}{$0.44\pm0.03$}     &\multicolumn{2}{c}{$0.9598\pm0.0007$} \\
        $\mathrm{P_{orb}[d]}$               & \multicolumn{2}{c}{$18.296\pm0.002$} & \multicolumn{2}{c}{$26.551\pm0.019$} & \multicolumn{2}{c}{$13.3364\pm0.0003$} \\
        $e$                                 & \multicolumn{2}{c}{$0.32\pm0.02$}    & \multicolumn{2}{c}{$0.371\pm0.003$}  & \multicolumn{2}{c}{$0.449\pm0.005$} \\
        \hline
    \end{tabular}
\end{table*}

\subsection{KIC 4930889}
KIC 4930889 was found to be an SB2 system and was characterised by \cite{papics2017} 
as consisting of a B5 IV-V primary and B8 IV-V secondary. Their orbital and spectral 
analysis placed both components in the iron-bump theoretical instability strip for 
g~modes in SPB stars. Their spectroscopic solution places the 
secondary as more evolved than the primary, which is an unphysical configuration 
unless binary evolution has altered the evolution of this system. We therefore revisit 
and re-normalised the original 26 spectra obtained by \cite{papics2017}, and derive a 
new spectroscopic solution, which is presented in Table \ref{tab:kepler_pars}. This new 
solution reports a much lower surface gravity for the primary and a much higher surface 
gravity for the secondary compared to the original solution reported by \cite{papics2017}. 
However the newly returned radii ratio ($\mathcal{R}=0.76$) is consistent with evolutionary 
expectations. Additionally, the effective temperatures of both component are lower by 
$\sim800 $ K compared to the solution of \cite{papics2017}. Seismic modelling of this 
system has not been performed so far.

\cite{papics2017} report 297 significant frequencies in the 4-year {\it Kepler} light 
curve after filtering for close peaks and low-order combinations. From this list of 297 
frequencies, the authors identify three separate period-spacing patterns \citep[Fig 15 
and 16 from][]{papics2017}. The first pattern consists of 20 consecutive radial orders and 
reveals a mean rotational frequency of $f_{\rm rot}=0.74\pm0.01 \, {\rm d^{-1}}$, 
following the method of \cite{vanReeth2016}. The slope of this pattern reveals it to 
consist of dipole prograde modes, leading to $\Pi_0=8712\pm320$ s. The second and third 
pattern are consistent with retrograde modes, but could not unambiguously be assigned a 
degree or component from which they originate. 

\begin{figure*}
    \centering
    \subfloat[\label{fig:kic4930889_single-MD}]{\includegraphics[width=0.45\textwidth]{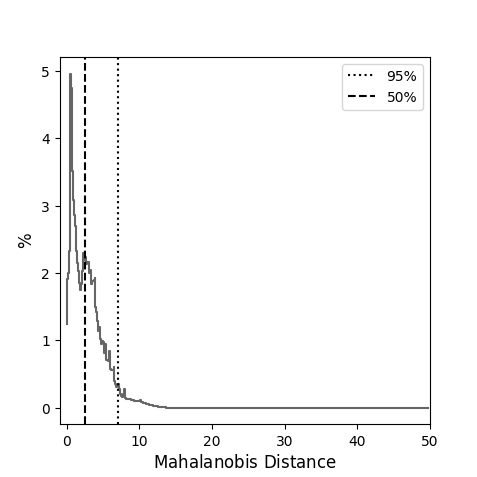}}
    ~
    \subfloat[\label{fig:kic4930889_single-corrs}]{\includegraphics[width=0.45\textwidth]{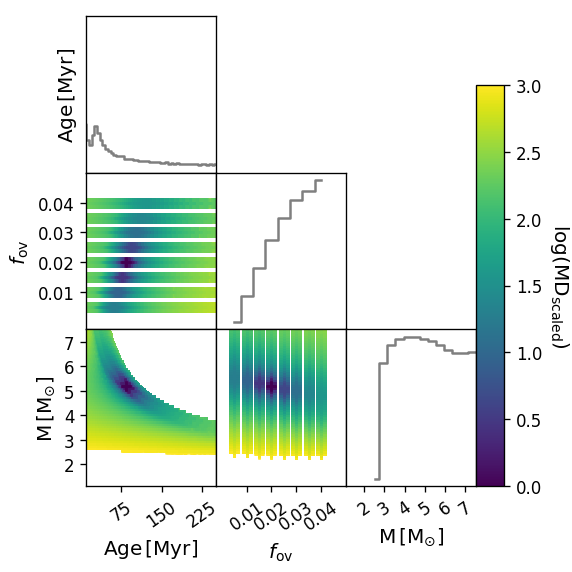}}
    \caption{\label{fig:kic4930889_single-all}Same as Fig. \ref{fig:hare_single-all} but for the single-star evaluation of KIC\,4930889\,A. }
\end{figure*}

\begin{figure*}
    \centering
    \subfloat[\label{fig:kic4930889_sb2-MD}]{\includegraphics[width=0.45\textwidth]{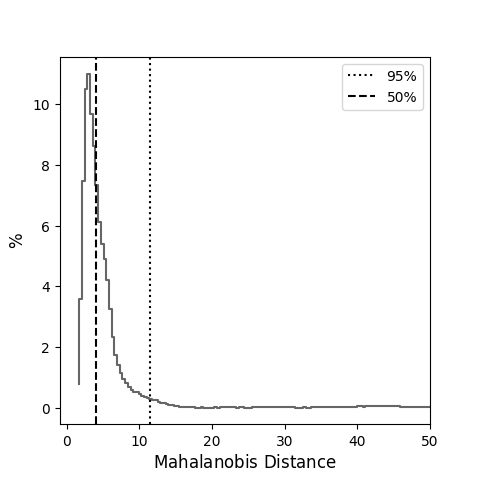}}
    ~
    \subfloat[\label{fig:kic4930889_sb2-corrs-primary}]{\includegraphics[width=0.45\textwidth]{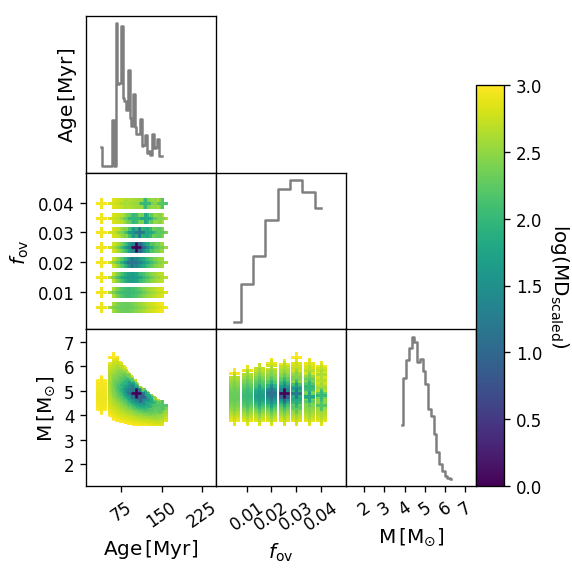}}
    \caption{\label{fig:kic4930889_sb2-all}Same as Fig. \ref{fig:hare_single-all} but for the SB2 evaluation of KIC\,4930889\,A. }
\end{figure*}

Figures \ref{fig:kic4930889_single-all} and \ref{fig:kic4930889_sb2-all} show the 
distributions of estimated parameters and their correlations. We find that the 
parameter estimates derived from the single star solution and the SB2 solution 
largely agree within their errors. While the best model returned by the binary 
evaluation is less massive by $0.28 \, {\rm M_{\odot}}$ compared to the single-star 
case, the convective core mass and location of the overshooting zone are have 
very good agreement between the single and binary evaluations. This is due to the fact 
that the seismic diagnostic $\Pi_0$ is strongly sensitive to the convective core 
mass. We note that the binary case greatly reduces both the mass-age and mass-overshoot 
degeneracies, which can be seen by comparing Figs \ref{fig:kic4930889_single-corrs} 
and \ref{fig:kic4930889_sb2-corrs-primary}. The binary constraints restrict the possible 
masses that can be considered valid at a given age, effectively lifting the degeneracy 
between the parameters. This propagates into the mass-overshoot degeneracy as the mass 
range is restricted. As the best solution, our analysis returns a system with age of 103 
Myr consisting of a ${\rm 4.89\substack{+1.49 \\ -1.09} \,\, M_{\odot}}$ primary with a 
${\rm 0.54} \, M_{\odot}$ convective core and a ${\rm 3.47\substack{+1.4 \\ -0.57} \,\, 
M_{\odot}}$ secondary with a ${\rm 0.60 \,\, M_{\odot}}$ convective core near the ZAMS. 
Finally, as an {\it a posteriori} check, the mass ratio of the best model agrees with 
the value listed in Table \ref{tab:kepler_pars} within $1\sigma$.

\subsection{KIC 6352430}
KIC 6352430 is a close ($\mathrm{P_{orb}=26.551\,d}$) eccentric ($e=0.37$) 
SB2 system consisting of a B7 V primary (KIC 6352430A) and F2.5 V secondary 
(KIC 6352430B) and was observed by \textit{Kepler} over 1459.5~d 
\citep{papics2013}. While some slight ellipsoidal variability was detected, 
the authors concluded that the remaining signal seen in the lightcurve could 
be explained by g modes excited via the $\kappa$-mechanism. Later analysis 
by \cite{papics2017} revealed 584 significant frequencies after cleaning for 
close peaks and low-order combination frequencies, from which a single sloped 
period-spacing pattern was identified consisting of 24 radial orders. The mean 
asymptotic period spacing and slope of the pattern identify it as SPB dipole 
prograde modes originating from the primary component, as much lower values 
are expected for the asymptotic dipole period spacing value in $\gamma$ Dor stars 
(\citealt{papics2017} vs. \citealt{vanReeth2015b}). Spectroscopic and orbital values 
taken from \cite{papics2013} are listed in Table \ref{tab:kepler_pars}. This 
system has not been the subject of modelling efforts to date. 

\begin{figure*}
    \centering
    \subfloat[\label{fig:kic6352430_single-MD}]{\includegraphics[width=0.45\textwidth]{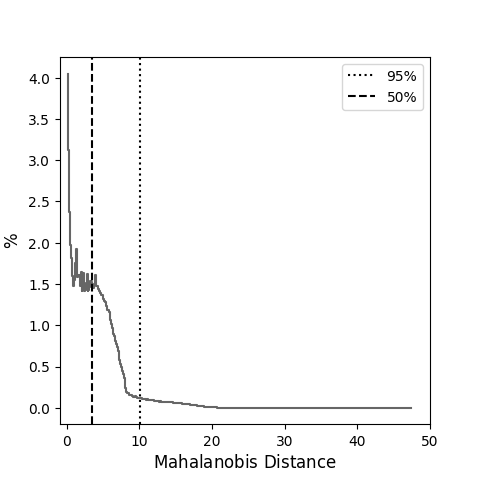}}
    ~
    \subfloat[\label{fig:kic6352430_single-corrs}]{\includegraphics[width=0.45\textwidth]{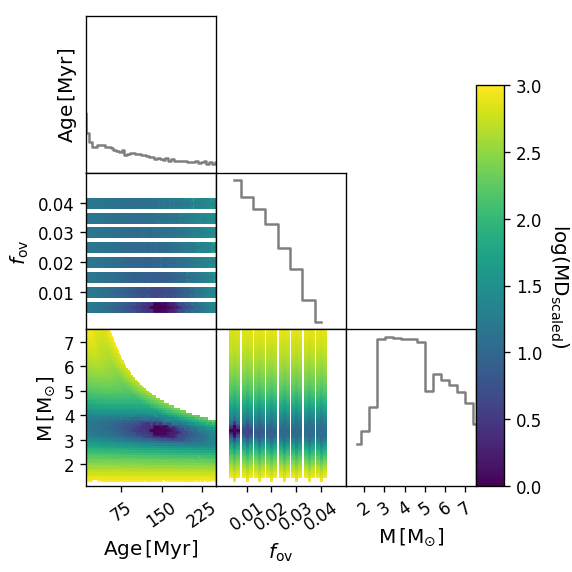}}
    \caption{\label{fig:kic6352430_single-all}Same as Fig. \ref{fig:hare_single-all} but for the single-star evaluation of KIC\,6352430\,A. }
\end{figure*}

\begin{figure*}
    \centering
    \subfloat[\label{fig:kic6352430_sb2-MD}]{\includegraphics[width=0.45\textwidth]{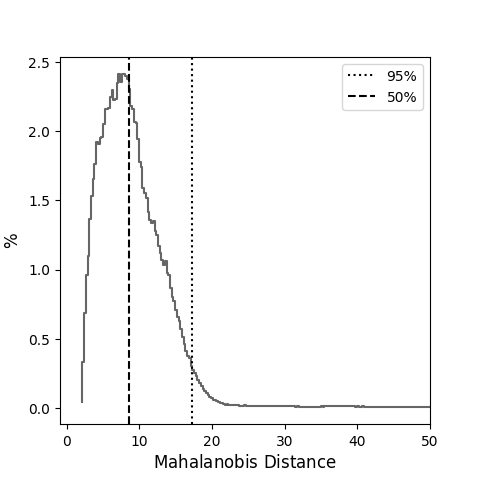}}
    ~
    \subfloat[\label{fig:kic6352430_sb2-corrs-primary}]{\includegraphics[width=0.45\textwidth]{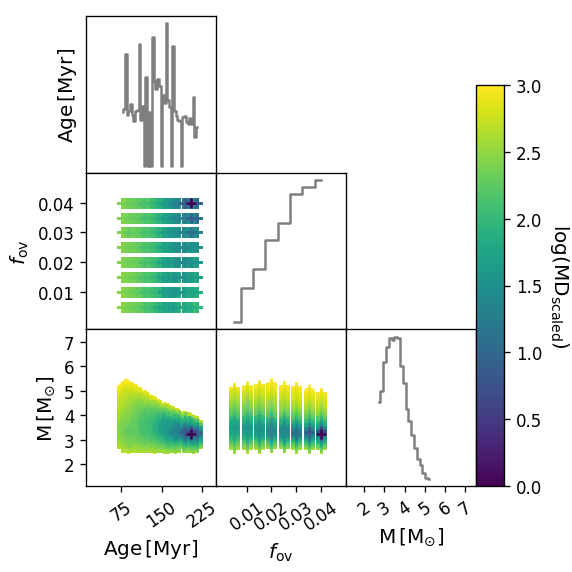}}
    \caption{\label{fig:kic6352430_sb2-all}Same as Fig. \ref{fig:hare_single-all} but for the SB2 evaluation of KIC\,6352430\,A. }
\end{figure*}

Figures \ref{fig:kic6352430_single-all} and \ref{fig:kic6352430_sb2-all} and 
Table \ref{tab:sample_soln} show the results for the modelling of KIC\,6352430. 
As in the case of KIC\,4930889, the estimates derived for the binary case are 
much more precise than that of the single-star case, and agree within the errors 
of the single-star solution. While the stellar mass and convective core mass 
agree across both the single-star and binary evaluation cases, the MLE 
errors show that we do not have the capacity to estimate overshooting. This suggests that 
the core mass rather than the extent and shape of overshooting is the 
important astrophysical quantity \citep{constantino2018}. We can see that the 
correlation between mass and overshoot that is present in the single-star 
correlation plots (Fig. \ref{fig:kic6352430_single-corrs}) has been effectively 
lifted in the binary case. The binary solution reveals a system with an age of 
205~Myr consisting of a ${\rm 3.23\substack{+1.98\\-0.56} \,\, M_{\odot}}$ primary 
with a ${\rm 0.56 \,\, M_{\odot}}$ convective core and a ${\rm 1.34\substack{+0.68\\-0.14} 
\, M_{\odot}}$ secondary with a ${\rm 0.08 \, M_{\odot}}$ convective core. The best 
estimated masses agree with the mass ratio reported in Table \ref{tab:kepler_pars} 
to within $1\sigma$.
\subsection{KIC 10080943}
KIC\,10080943 was the first system with two g-mode pulsating $\delta$ Sct/$\gamma$ Dor components, 
each of which were observed to have multiple pulsation patterns, for 
which binary modelling could be performed to obtain independent estimates 
for the component masses and radii from modelling the periastron brightening 
as KIC\,10080943 is an heartbeat star system \citep{keen2015,schmid2015,schmid2016}. 
In addition to identifying multiple patterns in each component, for both 
p- and g-modes, \cite{schmid2016} were able to derive surface-to-core 
rotation rate estimates and an independent age estimate from binary modelling. 
For our purposes, we take the values of ${\rm T_{eff}}$, $\log g$, 
${\rm M_{1,2}}$, and ${\rm R_{1,2}}$ from the binary modelling performed by 
\cite{schmid2015} and the estimates of $\Pi_0$ from the analysis of 
\cite{schmid2016}, all of which are listed in Table \ref{tab:kepler_pars}. 

Given the characterisation of this system, we can test how the application 
of different information in the modelling impacts our results. We
find that the single star and SB2 evaluation largely agree except for the 
overshoot and $X_{\rm c}$ returned by the best model for each case. 
The application of mass and radius estimates to the modelling procedure results 
in yet again different mass, age, and overshoot estimates for the best model. Again, 
we see that we have no capacity to provide MLE estimates of $f_{\rm ov}$. 
We find approximate agreement between ${\rm M_{cc}}$ for all cases. 

The difference in solution between the SB2 and heartbeat star cases is caused by the use
of the radii ratio in SB2 case versus individual radii in the heartbeat star case. Since 
KIC\,10080943 is comprised of two nearly identical stars, the application 
of the mass ratio and radii ratio does not provide any new information or strong constraints. Additionally, 
the covariance structure inherently changes between individual evolutionary tracks 
and the isoclouds, which produces the differences between the single star and SB2 
solutions. However, applying the absolute mass and radii estimates provides sufficient 
constraints to improve the solution. The best model from the heartbeat star evaluation reports a 
system with an age of $890\substack{+610\\-190}$ Myr consisting of a 
$1.99\substack{+0.37\\-0.49} \,\, {\rm M_{\odot}}$ primary with an 
$0.15 \, {\rm M_{\odot}}$ convective core and a $1.85\substack{+0.41\\-0.45} \,\, {\rm M_{\odot}}$ 
secondary with an $0.17 \, {\rm M_{\odot}}$ convective core. 

Of the six models that \cite{schmid2016} reported for KIC\,10080943, we are 
interested in comparing Models 4 and 6 to our result. Model 4 corresponds to 
the solution from modelling the individual g-mode periods for both components
accounting for the effect of rotation on pulsations using the Traditional 
Approximation of Rotation \citep[TAR][]{townsend2003,gyre2013}. Model 6 
corresponds to the solution from modelling the overall morphology of the g-mode 
period spacing pattern, while again accounting for rotation using the TAR. In 
both models, the diffusive exponential description of overshooting was used. 
\cite{schmid2016} only enforced an equal age constraint in their modelling and
employ a $\chi^2$ evaluation using only the individual g modes (per star) in 
model 4 and the g-mode period spacing pattern (per star) in model 6. Our single star
and SB2 solutions share approximate agreement with both model 4 and model 6 from 
\cite{schmid2016}. Our heartbeat star solution, which includes the mass ratio, absolute masses 
and radii, as well as the spectroscopic quantities, does not agree with either model 
4 nor model 6 from \cite{schmid2016}. Due to the differences in the modelling methodology, 
we suggest caution at making a direct comparison between the two results. 

\begin{figure*}
    \centering
    \subfloat[\label{fig:kic10080943_single-MD}]{\includegraphics[width=0.45\textwidth]{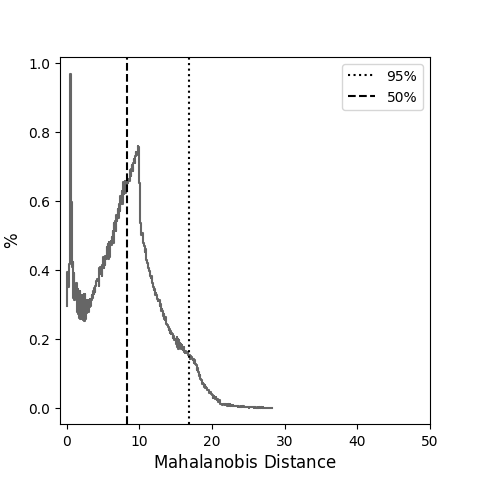}}
    ~
    \subfloat[\label{fig:kic10080943_single-corrs}]{\includegraphics[width=0.45\textwidth]{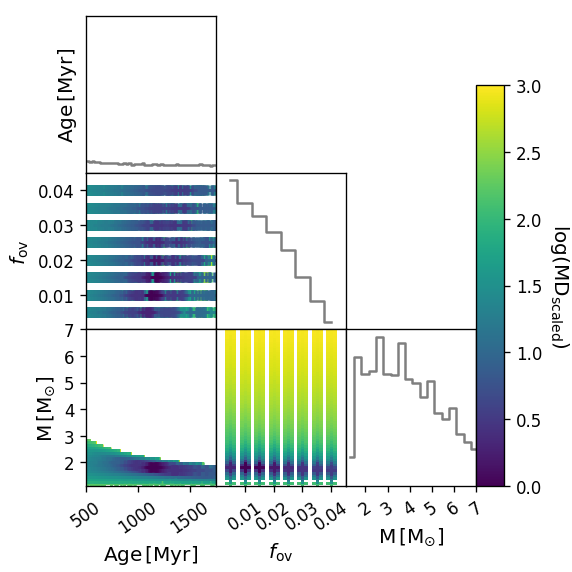}}
    \caption{\label{kic10080943_single-all} Same as Fig. \ref{fig:hare_single-all} but for the single-star evaluation of KIC\,10080943\,A. }
\end{figure*}

\begin{figure*}
    \centering
    \subfloat[\label{fig:kic10080943_sb2-MD}]{\includegraphics[width=0.45\textwidth]{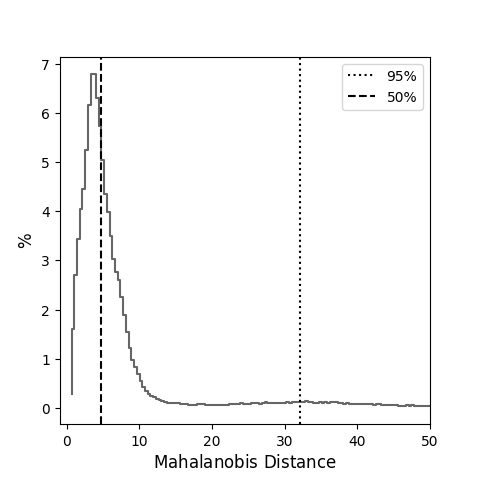}}
    ~
    \subfloat[\label{fig:kic10080943_sb2-corrs-primary}]{\includegraphics[width=0.45\textwidth]{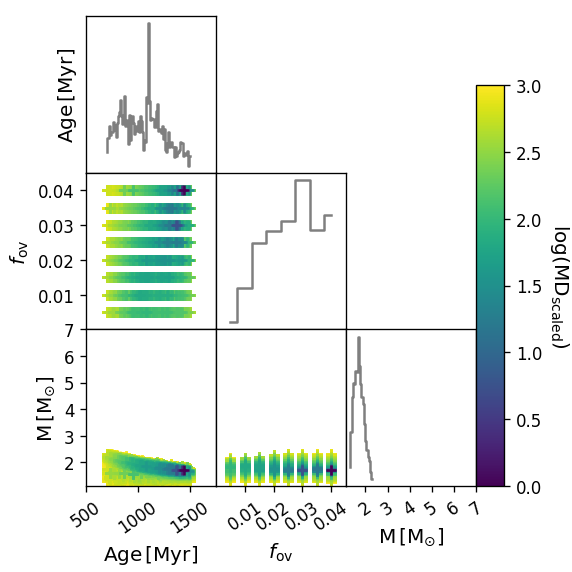}}
    \caption{\label{kic10080943_sb2-all}Same as Fig. \ref{fig:hare_single-all} but for the SB2 evaluation of KIC\,10080943\,A. }
\end{figure*}

\begin{figure*}
    \centering
    \subfloat[\label{fig:kic10080943_eb-MD}]{\includegraphics[width=0.45\textwidth]{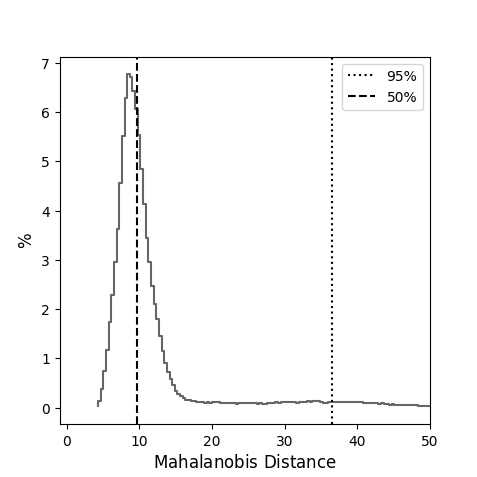}}
    ~
    \subfloat[\label{fig:kic10080943_eb-corrs-primary}]{\includegraphics[width=0.45\textwidth]{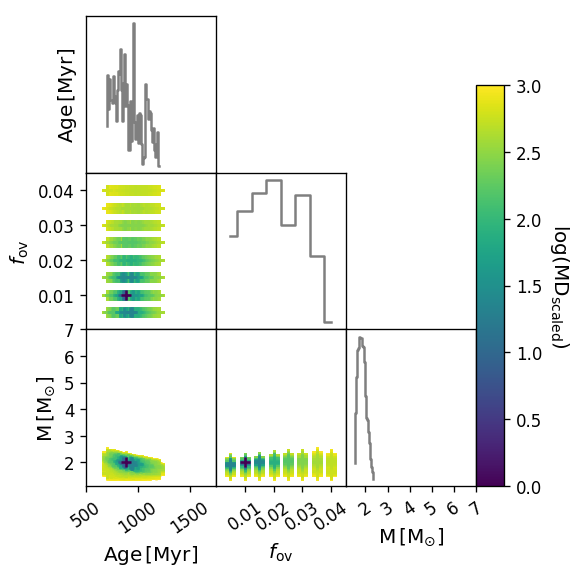}}
    \caption{\label{kic10080943_eb-all}Same as Fig. \ref{fig:hare_single-all} but for the heartbeat star evaluation of KIC\,10080943\,A. }
\end{figure*}

\begin{table*}
\tabcolsep=2pt
\centering
    \caption{ Maximum likelihood estimates (half inter-quartile range) of the model parameters for three {\it Kepler} binaries. Values reported with no lower and upper values are taken from the best model. 
    Top panel is for single star solution. Middle panel is for SB2 solution. Bottom panel is for heartbeat star solution where applicable.}
    \label{tab:sample_soln}
    \begin{tabular}{llcccccc}
        \hline
        Case & Parameter & \multicolumn{2}{c}{KIC 4930889} & \multicolumn{2}{c}{KIC 6352430} & \multicolumn{2}{c}{KIC 10080943} \\
        \hline
         \multirow{7}{*}{Single}       & ${\rm Age~[Myr]}$           & 85~(0,~2000)          & -- & 140~(0,~2000)         & -- & 1446  (0,  2500)      & -- \\
         &                               $f_{\rm ov}$                & 0.02  (0.005,  0.04)  & -- & 0.005~(--,~0.04)      & -- & 0.010  (0.005,  0.04) & -- \\
         &                               ${\rm M~[M_{\odot}]}$       & 5.2  (2.4,  7.0)      & -- & 3.4~(1.5,~8.9)        & -- & 1.7  (1.2,  9.7)      & -- \\
         &                               ${\rm R~[R_{\odot}]}$       & 6.39                  & -- & 2.91                  & -- & 2.66                  & -- \\
         &                               ${\rm M_{cc}~[M_{\odot}]}$  & 0.58                  & -- & 0.51                  & -- & 0.096                 & -- \\
         &                               ${\rm R_{ov}~[R_{\odot}]}$ & 0.37                  & -- & 0.31                  & -- & 0.103                 & -- \\
         &                               ${\rm X_{c}}$                 & 0.06                  & -- & 0.38                  & -- & 0.006                 & -- \\
        \hline
         \multirow{7}{*}{SB2} & ${\rm Age~[Myr]}$          & \multicolumn{2}{c}{103~(38,~150)}             & \multicolumn{2}{c}{205~(78,~215)}             & \multicolumn{2}{c}{1440~(700,~1500)}        \\
         &                      $f_{\rm ov}$               & 0.025~(0.005,~0.04)   & 0.005~(--,~0.04)      & 0.04~(0.005,~--)      & 0.005~(--,~0.04)      & 0.04~(0.005,~--)      & 0.04~(0.005,~--)    \\
         &                      ${\rm M~[M_{\odot}]}$      & 4.89~(3.80,~6.38)     & 3.47~(2.9,~4.87)      & 3.23~(2.67,~5.21)     & 1.34~(1.2,~2.02)      & 1.71~(1.26,~2.31)     & 1.56~(1.25,~2.21)   \\
         &                      ${\rm R~[R_{\odot}]}$      & 6.72                  & 2.65                  & 3.01                  & 1.42                  & 2.56                  & 1.79 \\
         &                      ${\rm M_{cc}~[M_{\odot}]}$ & 0.54                  & 0.60                  & 0.56                  & 0.08                  & 0.19                  & 0.18 \\
         &                      ${\rm R_{ov}~[R_{\odot}]}$ & 0.35                  & 0.33                  & 0.34                  & 0.11                  & 0.17                  & 0.17 \\
         &                      ${\rm X_{c}}$              & 0.05                  & 0.48                  & 0.42                  & 0.55                  & 0.28                  & 0.46 \\
        \hline
         \multirowcell{7}{heartbeat \\star} & ${\rm Age~[Myr]}$         & \multicolumn{2}{c}{--} & \multicolumn{2}{c}{--} & \multicolumn{2}{c}{890~(700,~1500)}        \\
                             & $f_{\rm ov}$               & -- & --                & -- & --                & 0.01~(0.005,~0.04)    & 0.005~(--,~0.04) \\
                             & ${\rm M~[M_{\odot}]}$      & -- & --                & -- & --                & 1.99~(1.50,~2.36)     & 1.85~(1.40,~2.26)    \\
                             & ${\rm R~[R_{\odot}]}$      & -- & --                & -- & --                & 3.11                  & 2.39  \\
                             & ${\rm M_{cc}~[M_{\odot}]}$ & -- & --                & -- & --                & 0.15                  & 0.17  \\
                             & ${\rm R_{ov}~[R_{\odot}]}$ & -- & --                & -- & --                & 0.14                  & 0.15  \\
                             & ${\rm X_{c}}$              & -- & --                & -- & --                & 0.06                  & 0.21 \\
        \hline
    \end{tabular}
\end{table*}

\section{Discussion}

Understanding the degeneracy between stellar mass, age, and extent of core 
overshooting is pivotal for asteroseismic modelling. The single star case 
evaluates $\Pi_0$, ${\rm T_{eff}}$, and $\log g$, all of which depend on 
a star's mass, age, and core mass. In the binary cases, our methodology 
imposes a strict range of ages at which a solution is valid. Since $\Pi_0$ 
varies with mass and age, by constraining the valid age range, we also 
constrain the masses at which a given $\Pi_0$ can be considered a valid 
solution. In the SB2 case, the mass ratio (when sufficiently far from unity)
drives the selection of stellar 
masses and core masses towards those combinations which satisfy the value 
of $\Pi_0$, which is already constrained by the valid age range. In the 
heartbeat star case, the addition of absolute masses and radii fixes the stellar 
masses to be considered in the valid age range, leaving only the extent 
of core overshooting to influence the mass of the core, and thus $\Pi_0$. 
While this is the most constrained case, the addition of the radii ratio 
in the SB2 case enables a cross-constraint on the evolution as well, 
constraining the mass-age-overshooting degeneracy to a manageable extent.

This work shows that the application of additional information derived from 
binarity significantly improves the results of extracted parameters and their uncertainties 
due to the independent cross constrains that binary information provides. In particular, 
it is seen that for each system that the single-star case has discrepant mass and age estimates 
compared to the SB2 case. Given the transformation of basis from age 
to mass for the isoclouds, the restricted age range imposed by the binary 
methodology corresponds to a restricted age range and thus shifts the best 
age and mass extracted by our methodology. This comparison of single to binary 
star solutions reveals the hierarchy of what results to take as robust, and 
which to reference with caution. The binary constraints render some single-star 
configurations impossible, allowing the precision of the seismic diagnostic to 
take full effect. We note that the inclusion of absolute mass and radius estimates 
only becomes important in the case where both stars are similar, with a mass ratio
near unity. In the case that the components of a binary are sufficiently different 
in mass, the inclusion of mass and radii ratios as well as the spectroscopic 
quantities of the secondary are sufficient to improve the model selection and parameter estimation. 

No clear trend emerges between mass and overshoot for this sample. Even accounting 
for the small sample, we do not encounter the same mass dependence of overshooting 
as seen by CT16, CT17, and CT18. Even with the addition of the asteroseismic diagnostic, 
we recover the entire input range of overshooting as the uncertainty on its estimates. 
However, we do find consistent estimates of the convective core mass and location of
the overshooting region \citep[c.f.][]{miglio2008,constantino2018}. Several overshoot 
values can satisfy the observations of a given target in the single-star seismic case, 
as seen in the correlation plots in Figs \ref{fig:kic4930889_single-corrs},
\ref{fig:kic6352430_single-corrs}, and \ref{fig:kic10080943_single-corrs}. We are able 
to remove this correlation structure by simultaneously applying asteroseismic and 
binary constraints, as seen in the correlation plots in Figs \ref{fig:kic4930889_sb2-corrs-primary}, 
\ref{fig:kic6352430_sb2-corrs-primary}, \ref{fig:kic10080943_sb2-corrs-primary}, and 
\ref{fig:kic10080943_eb-corrs-primary}. This leads to a unique determination of the 
stellar age and mass as seen by comparing the single and binary cases in Table 
\ref{tab:sample_soln}.

\section{Summary and Conclusions}
In this work we have formulated a framework for the modelling of g-mode pulsating stars 
in binary systems and applied it to three systems KIC\,4930889, KIC\,6352430, and KIC\,10080943 
\citep{papics2013,schmid2015,schmid2016,papics2017}. We calculated a grid of stellar 
evolution tracks using {\sc MESA} spanning a wide range of stellar masses and extents of 
core overshooting using the diffusive exponential overshooting description. We fixed 
$Z_{\rm ini}$, $\alpha_{\rm MLT}$, and $D_{\rm mix}$ after investigating the sensitivity 
of our observations to our grid of models. To carry out our modelling, we followed the 
approach discussed in detail by \cite{aerts2018b}.

To apply binary constraints and allow both components of a binary system to have a different 
amount of internal mixing, we introduced the concept of an iso(chrone-)cloud where the two 
components are only evaluated within the same isocloud. We modelled the three systems considering 
first the pulsating primary as single with the seismic diagnostic $\Pi_0$, then introduced 
the binary information in the form of age constraints, mass and radii ratios (KIC\,4930889, 
KIC\,6352430, KIC\,10080943), and absolute masses and radii (KIC 10080943).

The addition of binary information proved useful for reducing the uncertainties on parameter 
estimates and reducing the correlation between model parameters. Comparison of the MLE estimates 
derived by the MD calculations did not reveal any obvious dependence of overshooting with mass 
or age. Most interestingly, we did not recover the traditional binary mass discrepancy in our 
results. This is likely due to the fact that even in our most constrained case (KIC\,10080943) 
we do not have one\,per\,cent-level relative precision on the mass and radius estimates from binary
modelling. In the future, systems with the necessarily high precision on mass and radius estimates 
need to be scrutinised to determine if the mass discrepancy persists with the inclusion of asteroseismic 
information in the modelling procedure.

In this work, we do not include any possible effects of binary evolution or tidal effects on 
pulsations in our modelling. It has already been established for KIC\,10080943 that rotation 
has a much larger impact than the tidal interaction \citep{schmid2016}, and as both KIC\,4930889 
and KIC\,6352430 have longer orbital periods and smaller mass ratios, the tide generating 
potential in these systems is smaller than that of KIC\,10080943. The addition of more or 
longer period-spacing patterns in the determination of $\Pi_0$ can aid in the improvement 
of the relative precision of this parameter and would thus improve the derived parameters 
compared to the results of this work.

The methodology presented here can be extended to full modelling of g-mode period spacing patterns. 
Our isocloud evaluation methodology provides a robust framework for investigating the overall internal 
mixing in binary stars -- core overshooting as well as rotational or envelope mixing. Our methodology 
is also relevant for isochrone fitting of stellar populations such as clusters, where both $D_{\rm ov}$
and $D_{\rm min}$ should be allowed to vary from star to star. Application of this methodology to samples 
of EBs and clusters can provide insight into internal mixing phenomena at various ages and metallicities. 
Furthermore, the flexibility of this method allows for the easy inclusion of seismic information to impose 
an independent calibration of such phenomena, should it become available. Future modelling will investigate 
whether or not the core mass and extent of overshooting for single stars differs from those stars in binary 
systems, as this would indicate the impact of tidal forces and/or binary evolution. The recently launched NASA 
{\sc TESS} mission \citep{ricker2015} and the future ESA PLATO missions \citep{rauer2014} promise to deliver 
high-quality observations of thousands of new g-mode pulsators with sufficiently high precision on $\Pi_0$ 
for stars in binaries and clusters (only in the continuous viewing zones for TESS), all of which will 
be suitable for analysis under the methodology put forth here. After their future release (2021+), the addition
of Gaia astrometric binary solutions for non-eclipsing binary systems will enable the 
application of direct mass estimates in this methodology, instead of using only the mass ratios as was done for the 
SB2 systems here \citep{lindegren2018}.

\section*{Acknowledgements}
The authors thank the referee for his/her/their comments and suggestions towards improving the manuscript.
The research leading to these results has received funding from the European Research Council (ERC) under the 
European Union's Horizon 2020 research and innovation programme (grant agreement N$^\circ$670519: MAMSIE). The 
computational resources and services used in this work were provided by the VSC (Flemish Supercomputer Center), 
funded by the Research Foundation - Flanders (FWO) and the Flemish Government – department EWI.


 

\bibliographystyle{mnras}
\bibliography{johnston_submitted_v2.bib}


\appendix


\section{Secondary component parameter correlation plots}
\begin{figure}
    \includegraphics[width=0.45\textwidth]{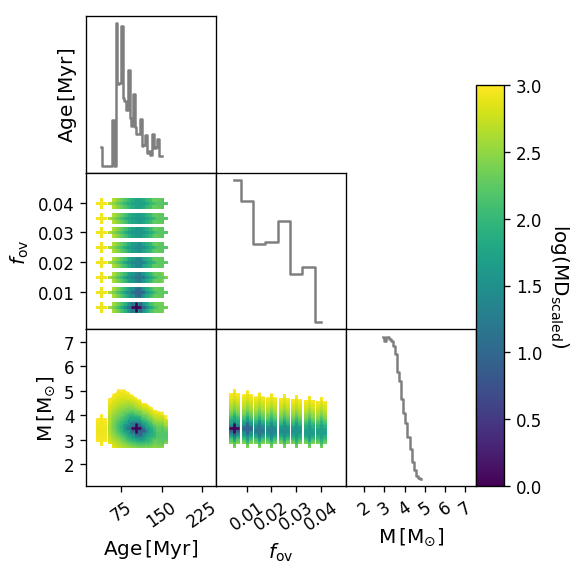}
    \caption{Same as Fig. \ref{fig:hare_single-corrs} but for the SB2 evaluation of KIC\,4930889\,B.}
    \label{fig:kic4930889_sb2-corrs-secondary}
\end{figure}

\begin{figure}
    \includegraphics[width=0.45\textwidth]{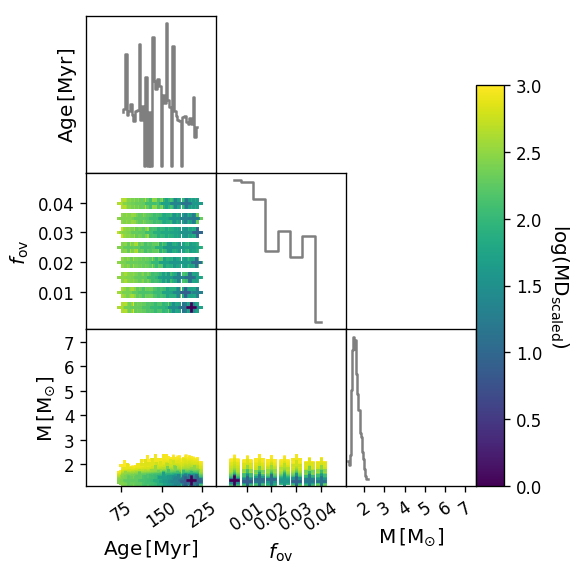}
	\caption{Same as Fig. \ref{fig:hare_single-corrs} but for the SB2 evaluation of KIC\,6352430\,B.}
    \label{fig:kic6352430_sb2-corrs-secondary}
\end{figure}

\begin{figure}
	\includegraphics[width=0.45\textwidth]{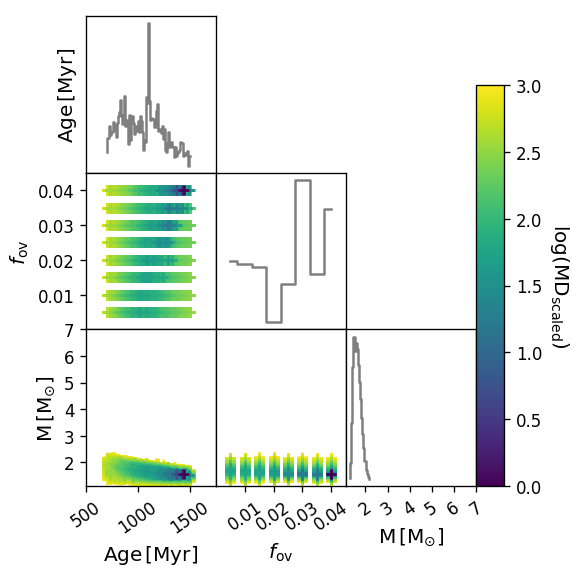}
    \caption{Same as Fig. \ref{fig:hare_single-corrs} but for the SB2 evaluation of KIC\,10080943\,B.}
    \label{fig:kic10080943_sb2-corrs-secondary}
\end{figure}
\begin{figure}
	\includegraphics[width=0.45\textwidth]{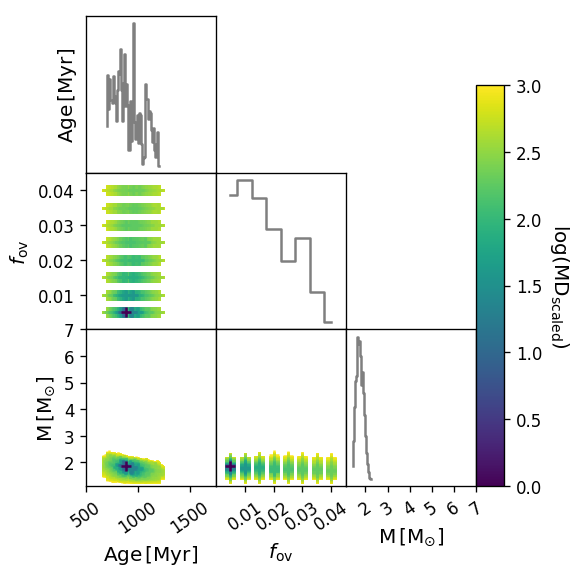}
    \caption{Same as Fig. \ref{fig:hare_single-corrs} but for the heartbeat star evaluation of KIC\,10080943\,B.}
    \label{fig:kic10080943_eb-corrs-secondary}
\end{figure}


\bsp	
\label{lastpage}

\end{document}